\documentclass[a4paper]{jpconf}
\usepackage{graphicx}
\usepackage[hidelinks]{hyperref}
\usepackage{color}
\usepackage{amsmath,amsfonts,amssymb,bm}

\usepackage{lipsum}

\usepackage{fancyhdr}
\def\beq{\begin{equation}}
\def\eeq{\end{equation}}
\def\beqn{\begin{eqnarray}}
\def\eeqn{\end{eqnarray}}
\def\beqs{\begin{subequations}}
\def\eeqs{\end{subequations}}

\def\P {{\bf P}}

\def\bi {\mbox{\boldmath $i$}}

\def\P {{\bf P}}

\begin{document}

\title{Fragmentation of a trapped bosonic mixture}

\author{O~E~Alon$^{1,2}$ and L~S~Cederbaum$^{3}$}
\address{$^{1}$ Department of Physics, University of Haifa, 3498838 Haifa, Israel}
\address{$^{2}$ Haifa Research Center for Theoretical Physics and Astrophysics, University of Haifa, 3498838 Haifa, Israel}
\address{$^{3}$ Theoretical Chemistry, Physical Chemistry Institute, Heidelberg University, D-69120 Heidelberg, Germany}

\ead{ofir@research.haifa.ac.il}

\begin{abstract}
Fragmentation of bosons and pairs in a trapped imbalanced bosonic mixture is investigated analytically using an
exactly solvable model, the generic harmonic-interaction model for mixtures.
Closed-form expressions for the eigenvalues and eigenfunctions of the reduced one-particle and two-particle
density matrices as a function of all parameters,
the masses, numbers of bosons, and the intraspecies and interspecies interactions,
are obtained and analyzed.
As an application, we consider a system made of $N_1=100$ non-interacting species $1$ bosons embedded in a bath
made of $N_2=10^6$ non-interacting species $2$ bosons,
and show how fragmentation of the system's bosons and pairs emerges from
the system--bath interaction only.
Interestingly, the lighter the bosons comprising the bath are the stronger is the system's fragmentation.
Further applications are briefly discussed.    
\end{abstract}

\section{Introduction}\label{INTRODUCTION}

Fragmentation of Bose-Einstein condensates has long and continuously drawn much interest
[1-22].
Fragmentation implies the macroscopic occupation of more than a single natural orbital
of the reduced one-particle density matrix \cite{Lowdin_1955,RDM_2000}, the reduced two-particle density matrix \cite{RDM_2008},
and so on.
For many years, bosonic mixtures have been a vast and fertilized ground for studying many-body phenomena
[26-54].
Fragmentation in bosonic mixtures can obviously be richer but is generally less studied.
For instance, how one species of bosons might induce fragmentation on a second species,
just by interacting with it, is intriguing to explore.

Let us address the tool of choice.
Generally, many-particle systems can be explored numerically or analytically.
In the present work we follow the latter.
Harmonic-interaction models have been used extensively to explore many-particle systems,
including Bose-Einstein condensates, see, e.g., 
[55-75].
Here, we use the generic harmonic-interaction model for bosonic mixtures \cite{IJQC_1991,EPJD_2014,JPA_2017}
to derive the eigenvalues and eigenfunctions of the reduced one-particle and two-particle
density matrices,
and investigate fragmentation in an imbalanced bosonic mixture.
The present investigation and results generalize
some recent outcomes obtained for the specific case of
a symmetric, i.e., balanced, bosonic mixture \cite{Atoms_2021}.

\section{Theory}\label{THEORY}

The generic harmonic-interaction model for bosonic mixtures reads \cite{IJQC_1991,EPJD_2014,JPA_2017}
\beqn\label{HAM_HIM_MIX}
& & \hat H(x_1,\ldots,x_{N_1},y_1,\ldots,y_{N_2}) = \nonumber \\
& & = \sum_{j=1}^{N_1} \left( -\frac{1}{2m_1} \frac{\partial^2}{\partial x_j^2} + \frac{1}{2} m_1\omega^2 x_j^2 \right)  
+ \sum_{j=1}^{N_2} \left( -\frac{1}{2m_2} \frac{\partial^2}{\partial y_j^2} +
\frac{1}{2} m_2\omega^2 y_j^2 \right) + \nonumber \\
& & + 
\lambda_1 \sum_{1 \le j < k}^{N_1} \left(x_j-x_k\right)^2 + 
\lambda_2 \sum_{1 \le j < k}^{N_2} \left(y_j-y_k\right)^2 +  
\lambda_{12} \sum_{j=1}^{N_1} \sum_{k=1}^{N_2} \left(x_j-y_k\right)^2. \
\eeqn
Here, $m_1$ and $m_2$ are the masses, $N_1$ and $N_2$ the numbers of species $1$ and species $2$
bosons, $\lambda_1$ and $\lambda_2$ the intraspecies interactions,
and $\lambda_{12}$ the interspecies interaction.
$\omega$ is the frequency of the trapping potential and $\hbar=1$ throughout this work.
For simplicity and without loss of generality we work in one spatial dimension.

The all-particle density matrix associated
with the ground state of (\ref{HAM_HIM_MIX}) has been
obtained in \cite{JPA_2017} and reads 
\beqn\label{N_DENS}
& & \!\!\!\!\!\!\!\!
\rho^{(N)}(x_1,\ldots,x_{N_1},y_1,\ldots,y_{N_2},x'_1,\ldots,x'_{N_1},y'_1,\ldots,y'_{N_2}) = \nonumber \\
& & \!\!\!\!\!\!\!\!
= \left(\frac{m_1\Omega_1}{\pi}\right)^{\frac{N_1-1}{2}}
\left(\frac{m_2\Omega_2}{\pi}\right)^{\frac{N_2-1}{2}}
\left(\frac{M_{12}\Omega_{12}}{\pi}\right)^{\frac{1}{2}}
\left(\frac{M\omega}{\pi}\right)^{\frac{1}{2}} \times \nonumber \\
& & \!\!\!\!\!\!\!\!
\times e^{-\frac{\alpha_1}{2} \sum_{j=1}^{N_1} \left(x_j^2+{x'_j}^2\right) - 
\beta_1 \sum_{1 \le j < k}^{N_1} \left(x_j x_k + x'_j x'_k\right)}
e^{-\frac{\alpha_2}{2} \sum_{j=1}^{N_2} \left(y_j^2+{y'_j}^2\right) - 
\beta_2 \sum_{1 \le j < k}^{N_2} \left(y_j y_k + y'_j y'_k\right)} \times \nonumber \\
& & \!\!\!\!\!\!\!\!
\times e^{+\gamma \sum_{j=1}^{N_1} \sum_{k=1}^{N_2} \left(x_j y_k + x'_j y'_k\right)},  \
\eeqn 
where
$\Omega_{1} = \sqrt{\omega^2 + \frac{2}{m_1}\left(N_1\lambda_1+N_2\lambda_{12}\right)}$,
$\Omega_{2} = \sqrt{\omega^2 + \frac{2}{m_2}\left(N_2\lambda_2+N_1\lambda_{12}\right)}$,
and
$\Omega_{12} = \sqrt{\omega^2 + 2\left(\frac{N_1}{m_2}+\frac{N_2}{m_1}\right)\lambda_{12}}$
are the mixture's frequencies and
$M=m_1N_1+m_2N_2$ and $M_{12}=\frac{m_1m_2}{M}$
its masses.
There are three kinds of terms,
one-particle terms with parameters $\alpha_1$ and $\alpha_2$,
two-particle intraspecies terms with parameters $\beta_1$ and $\beta_2$,
and
the two-particle interspecies term with parameter $\gamma$.
These parameters govern the properties of the mixture \cite{JPA_2017}, also see below,
and are collected for compactness in (\ref{MIX_WF_DENS_PAR}) of the appendix.

To obtain the reduced one-particle density matrices from (\ref{N_DENS}) one must integrate out
all $N_1$ species $1$ coordinates and $N_2-1$ species $2$ coordinates or
vice versa.
The final result is \cite{JPA_2017}
\beqn\label{SPECIES_1_and_2_1RDM}
& &
\rho^{(1)}_1(x,x')
= N_1 \left(\frac{\alpha_1+C_{1,0}}{\pi}\right)^{\frac{1}{2}}
e^{-\frac{\alpha_1+\frac{C_{1,0}}{2}}{2}\left(x^2+{x'}^2\right)} 
e^{- \frac{1}{2} C_{1,0} xx'}, \nonumber \\
& &
\rho^{(1)}_2(y,y')
= N_2 \left(\frac{\alpha_2+C'_{0,1}}{\pi}\right)^{\frac{1}{2}}
e^{-\frac{\alpha_2+\frac{C'_{0,1}}{2}}{2}\left(y^2+{y'}^2\right)} 
e^{- \frac{1}{2} C'_{0,1} yy'}. \
\eeqn
The coefficients $C_{1,0}$ and $C'_{0,1}$ are a result
of recursive relations and depend on the all-particle density's parameters
$\alpha_1$, $\alpha_2$, $\beta_1$, $\beta_2$, and $\gamma$.
The working expressions are collected in (\ref{SPECIES1_12},\ref{SPECIES2_12}).
This is our starting point.

To prescribe the eigenvalues and eigenfunctions of (\ref{SPECIES_1_and_2_1RDM})
and be able to analyze them analytically,
we must work out simplified expressions for $C_{1,0}$ and $C'_{0,1}$.
Plugging (\ref{MIX_WF_DENS_PAR}) into (\ref{SPECIES1_12},\ref{SPECIES2_12})
one readily finds
\beqn\label{SPECIES_1_and_2_1RDM_COEFFS}
& &
\alpha_1 + C_{1,0} = m_1 \Omega_1 \frac{1}{1 + \frac{1}{N_1} \left( \frac{m_1N_1}{M} \frac{\Omega_1}{\omega} +
\frac{m_2N_2}{M} \frac{\Omega_1}{\Omega_{12}} - 1 \right)}, \nonumber \\
& &
\alpha_2 + C'_{0,1} = m_2 \Omega_2 \frac{1}{1 + \frac{1}{N_2} \left( \frac{m_2N_2}{M} \frac{\Omega_2}{\omega} +
\frac{m_1N_1}{M} \frac{\Omega_2}{\Omega_{12}} - 1 \right)}. \
\eeqn
Together with the $\alpha_1$ and $\alpha_2$ parameters,
these will govern the degree of bosons' fragmentation of the mixture.

Equation (\ref{SPECIES_1_and_2_1RDM}) implies a straightforward application of Mehler's formula,
see \cite{Robinson_1977,Schilling_2013,Atoms_2021},
which reads
$\left[\frac{(1-\rho)s}{(1+\rho)\pi}\right]^{\frac{1}{2}}
e^{-\frac{1}{2}\frac{(1+\rho^2)s}{1-\rho^2}\left(x^2+{x'}^2\right)}e^{+\frac{2\rho s}{1-\rho^2}xx'}
= \sum_{n=0}^\infty (1-\rho)\rho^n \frac{1}{\sqrt{2^n n!}} \left(\frac{s}{\pi}\right)^{\frac{1}{4}} H_n(\sqrt{s}x) e^{-\frac{1}{2}s x^2} \times$
$\times \frac{1}{\sqrt{2^n n!}} \left(\frac{s}{\pi}\right)^{\frac{1}{4}} H_n(\sqrt{s}x') e^{-\frac{1}{2}s {x'}^2}$,
where $s>0$ is the scaling,
$1 > \rho \ge 0$,
and
$H_n$ are the Hermite polynomials.
By comparing the respective structures,
one determines
the condensate fractions
of species $1$ and of species $2$
\beqn\label{1RDMs_COND_FRAC_1_2}
& & 1 - \rho^{(1)}_1 = \frac{2\sqrt{\alpha_1+C_{1,0}}}{\sqrt{\alpha_1} + \sqrt{\alpha_1+C_{1,0}}} = \nonumber \\
& & = \frac{2}{\sqrt{\left[1 + \frac{1}{N_1} \left( \frac{m_1N_1}{M} \frac{\omega}{\Omega_1} + \frac{m_2N_2}{M} \frac{\Omega_{12}}{\Omega_1} - 1 \right)\right]\left[1 + \frac{1}{N_1} \left( \frac{m_1N_1}{M} \frac{\Omega_1}{\omega} +
\frac{m_2N_2}{M} \frac{\Omega_1}{\Omega_{12}} - 1 \right)\right]}+1}, \nonumber \\
& & 1 - \rho^{(1)}_2 = \frac{2\sqrt{\alpha_2+C'_{0,1}}}{\sqrt{\alpha_2} + \sqrt{\alpha_2+C'_{0,1}}} = \nonumber \\
& & = \frac{2}{\sqrt{\left[1 + \frac{1}{N_2} \left( \frac{m_2N_2}{M} \frac{\omega}{\Omega_2} + \frac{m_1N_1}{M} \frac{\Omega_{12}}{\Omega_2} - 1 \right)\right]\left[1 + \frac{1}{N_2} \left( \frac{m_2N_2}{M} \frac{\Omega_2}{\omega} +
\frac{m_1N_1}{M} \frac{\Omega_2}{\Omega_{12}} - 1 \right)\right]}+1}. \
\eeqn
The corresponding depleted fractions are
\beqn\label{1RDMs_DEPLT_FRAG_1_2}
& & \rho^{(1)}_1 = \frac{\sqrt{\alpha_1} - \sqrt{\alpha_1+C_{1,0}}}{\sqrt{\alpha_1} + \sqrt{\alpha_1+C_{1,0}}} = \nonumber \\
& & = \frac{\sqrt{\left[1 + \frac{1}{N_1} \left( \frac{m_1N_1}{M} \frac{\omega}{\Omega_1} + \frac{m_2N_2}{M} \frac{\Omega_{12}}{\Omega_1} - 1 \right)\right]\left[1 + \frac{1}{N_1} \left( \frac{m_1N_1}{M} \frac{\Omega_1}{\omega} +
\frac{m_2N_2}{M} \frac{\Omega_1}{\Omega_{12}} - 1 \right)\right]}-1}{\sqrt{\left[1 + \frac{1}{N_1} \left( \frac{m_1N_1}{M} \frac{\omega}{\Omega_1} + \frac{m_2N_2}{M} \frac{\Omega_{12}}{\Omega_1} - 1 \right)\right]\left[1 + \frac{1}{N_1} \left( \frac{m_1N_1}{M} \frac{\Omega_1}{\omega} +
\frac{m_2N_2}{M} \frac{\Omega_1}{\Omega_{12}} - 1 \right)\right]}+1}, \nonumber \\
& & \rho^{(1)}_2 = \frac{\sqrt{\alpha_2} - \sqrt{\alpha_2+C'_{0,1}}}{\sqrt{\alpha_2} + \sqrt{\alpha_2+C'_{0,1}}} = \nonumber \\
& & = \frac{\sqrt{\left[1 + \frac{1}{N_2} \left( \frac{m_2N_2}{M} \frac{\omega}{\Omega_2} + \frac{m_1N_1}{M} \frac{\Omega_{12}}{\Omega_2} - 1 \right)\right]\left[1 + \frac{1}{N_2} \left( \frac{m_2N_2}{M} \frac{\Omega_2}{\omega} +
\frac{m_1N_1}{M} \frac{\Omega_2}{\Omega_{12}} - 1 \right)\right]}-1}{\sqrt{\left[1 + \frac{1}{N_2} \left( \frac{m_2N_2}{M} \frac{\omega}{\Omega_2} + \frac{m_1N_1}{M} \frac{\Omega_{12}}{\Omega_2} - 1 \right)\right]\left[1 + \frac{1}{N_2} \left( \frac{m_2N_2}{M} \frac{\Omega_2}{\omega} +
\frac{m_1N_1}{M} \frac{\Omega_2}{\Omega_{12}} - 1 \right)\right]}+1}. \
\eeqn
The latter dictate the degree of fragmentation of each species.
Furthermore,
the decomposition of the mixture's one-particle reduced
density matrices (\ref{SPECIES_1_and_2_1RDM}) 
is simply
\beqn\label{SPECIES1_SPECIES2_1_RDM_NAT_ORB}
& &
\rho_1^{(1)}(x,x') = 
N_1 \sum_{n_1=0}^\infty \left(1-\rho_1^{(1)}\right)\left(\rho_1^{(1)}\right)^{n_1}
\Phi^{(1)}_{1,{n_1}}(x) \Phi^{(1),\ast}_{1,{n_1}}(x'), \nonumber \\
& & \Phi^{(1)}_{1,n_1}(x) =
\frac{1}{\sqrt{2^{n_1} {n_1}!}}
\left(\frac{s_1^{(1)}}{\pi}\right)^{\frac{1}{4}}
H_{n_1}\left(\sqrt{s_1^{(1)}}x\right)
e^{-\frac{1}{2}s_1^{(1)}x^2}, \nonumber \\
& &
\rho_2^{(2)}(y,y') = 
N_2 \sum_{n_2=0}^\infty \left(1-\rho_2^{(1)}\right)\left(\rho_2^{(1)}\right)^{n_2}
\Phi^{(1)}_{2,{n_2}}(y) \Phi^{(1),\ast}_{2,{n_2}}(y'), \nonumber \\
& & \Phi^{(1)}_{2,{n_2}}(y) =
\frac{1}{\sqrt{2^{n_2} {n_2}!}}
\left(\frac{s_2^{(1)}}{\pi}\right)^{\frac{1}{4}}
H_{n_2}\left(\sqrt{s_2^{(1)}}y\right)
e^{-\frac{1}{2}s_2^{(1)}y^2}, \
\eeqn
where the scalings are
\beqn\label{SPECIES1_SPECIES2_1_RDM_NAT_ORB_SCALINGS}
& & s^{(1)}_1 = \sqrt{\alpha_1\left(\alpha_1+C_{1,0}\right)} =
m_1\Omega_1
\sqrt{\frac{1 + \frac{1}{N_1} \left( \frac{m_1N_1}{M} \frac{\omega}{\Omega_1} + \frac{m_2N_2}{M} \frac{\Omega_{12}}{\Omega_1} - 1 \right)}{1 + \frac{1}{N_1} \left( \frac{m_1N_1}{M} \frac{\Omega_1}{\omega} +
\frac{m_2N_2}{M} \frac{\Omega_1}{\Omega_{12}} - 1 \right)}}, \nonumber \\
& & s^{(1)}_2 = \sqrt{\alpha_2\left(\alpha_2+C'_{0,1}\right)} =
m_2\Omega_2
\sqrt{\frac{1 + \frac{1}{N_2} \left( \frac{m_2N_2}{M} \frac{\omega}{\Omega_2} + \frac{m_1N_1}{M} \frac{\Omega_{12}}{\Omega_2} - 1 \right)}{1 + \frac{1}{N_2} \left( \frac{m_2N_2}{M} \frac{\Omega_2}{\omega} +
\frac{m_1N_1}{M} \frac{\Omega_2}{\Omega_{12}} - 1 \right)}}. \
\eeqn
Hence, we obtain in closed form the
condensate fractions, depleted fractions, and 
the natural orbitals of
species $1$ and $2$ bosons as a function of all parameters of the mixture.
Because of the structure of the eigenvalues, see (\ref{SPECIES1_SPECIES2_1_RDM_NAT_ORB}), 
the depletions $\rho_1^{(1)}$ and $\rho_2^{(1)}$ determine the degree of fragmentation of each species.
Consider, e.g., $\rho_1^{(1)}=\frac{1}{2}$.
Then, fragmentation of species $1$ takes place with the first few natural occupation numbers being
$50\%$, $25\%$, $12.5\%$, $6.25\%$, $3.125\%$ and so on.
Increasing the depletion, for instance to $\rho_1^{(1)}=\frac{1}{\sqrt{2}}$,
the fragmentation becomes stronger and the first few natural occupation numbers become
$29.29\%$, $20.71\%$, $14.64\%$, $10.36\%$, $7.322\%$ and so on.
Finally, because the mixture is imbalanced
one generally finds different degrees of fragmentation for each of the species
and can investigate how fragmentation depends on all mixture's parameters,
see the illustrative example below.

The reduced two-particle density matrices are obtained by integrating out
all $N_1$ species $1$ coordinates and $N_2-2$ species $2$ coordinates or
vice versa from (\ref{N_DENS}).
The final result is \cite{JPA_2017,Floquet_2020}
\beqn\label{SPECIES_1_and_2_2RDM}
& &
\rho_1^{(2)}(x_1,x_2,x'_1,x'_2) = 
N_1(N_1-1) \left(\frac{\alpha_1+C_{1,0}}{\pi}\right)^{\frac{1}{2}} \left(\frac{\alpha_1+C_{2,0}}{\pi}\right)^{\frac{1}{2}} \times \nonumber \\
& & 
\times e^{-\frac{\alpha_1}{2}\left(x_1^2+x_2^2+{x'_1}^2+{x'_2}^2\right)} e^{-\beta_1\left(x_1x_2 + x'_1x'_2\right)}
e^{-\frac{1}{4} C_{2,0} \left(x_1+x_2+x'_1+x'_2\right)^2}, \nonumber \\
& &
\rho_2^{(2)}(y_1,y_2,y'_1,y'_2) = 
N_2(N_2-1) \left(\frac{\alpha_2+C'_{0,1}}{\pi}\right)^{\frac{1}{2}} \left(\frac{\alpha_2+C'_{0,2}}{\pi}\right)^{\frac{1}{2}} \times \nonumber \\
& & 
\times e^{-\frac{\alpha_2}{2}\left(y_1^2+y_2^2+{y'_1}^2+{y'_2}^2\right)} e^{-\beta_2\left(y_1y_2 + y'_1y'_2\right)}
e^{-\frac{1}{4} C'_{0,2} \left(y_1+y_2+y'_1+y'_2\right)^2}. \
\eeqn
The coefficients $C_{2,0}$ and $C'_{0,2}$ are 
the predecessors of the coefficients $C_{1,0}$ and $C'_{0,1}$ in the recursive relations 
resulting when reducing the all-particle density matrix (\ref{N_DENS})
and are functions of
$\alpha_1$, $\alpha_2$, $\beta_1$, $\beta_2$, and $\gamma$,
see (\ref{SPECIES1_12},\ref{SPECIES2_12}).
Our treatment of the natural pair functions of (\ref{SPECIES_1_and_2_2RDM}) and their occupations
in the imbalanced mixture begins here.

The first step
is to work with the appropriate set of coordinates.
There are four coordinates in each of the expressions (\ref{SPECIES_1_and_2_2RDM}),
suggesting that one should work with pairs.
Defining the center-of-mass $\frac{x_1+x_2}{2}$ and relative $x_1-x_2$
coordinates of a pair of species $1$ bosons, and similarly for species $2$ bosons,
the reduced two-particle density matrices transform to
\beqn\label{SPECIES1_SPECIES2_2_RDM_DIAG}
& &
\!\!\!\!\!\!\!\!
\rho_1^{(2)}(x_1,x'_1,x_2,x'_2) = 
N_1(N_1-1)
\left(\frac{\alpha_1-\beta_1}{2\pi}\right)^{\frac{1}{2}}
e^{-\frac{\frac{1}{2}\left(\alpha_1-\beta_1\right)}{2}\left[\left(x_1-x_2\right)^2+\left(x'_1-x'_2\right)^2\right]} \times \nonumber \\
& &
\!\!\!\!\!\!\!\!
\times
\left(\frac{2\left[\alpha_1+\beta_1+2C_{2,0}\right]}{\pi}\right)^{\frac{1}{2}}
e^{-\frac{2\left[\alpha_1+\beta_1+C_{2,0}\right]}{2}\left[\left(\frac{x_1+x_2}{2}\right)^2+\left(\frac{x'_1+x'_2}{2}\right)^2\right]}
e^{-2C_{2,0} \left(\frac{x_1+x_2}{2}\right)\left(\frac{x'_1+x'_2}{2}\right)}, \nonumber \\
& &
\!\!\!\!\!\!\!\!
\rho_2^{(2)}(y_1,y'_1,y_2,y'_2) = 
N_2(N_2-1)
\left(\frac{\alpha_2-\beta_2}{2\pi}\right)^{\frac{1}{2}}
e^{-\frac{\frac{1}{2}\left(\alpha_2-\beta_2\right)}{2}\left[\left(y_1-y_2\right)^2+\left(y'_1-y'_2\right)^2\right]} \times \nonumber \\
& &
\!\!\!\!\!\!\!\!
\times
\left(\frac{2\left[\alpha_2+\beta_2+2C'_{0,2}\right]}{\pi}\right)^{\frac{1}{2}}
e^{-\frac{2\left[\alpha_2+\beta_2+C'_{0,2}\right]}{2}\left[\left(\frac{y_1+y_2}{2}\right)^2+\left(\frac{y'_1+y'_2}{2}\right)^2\right]}
e^{-2C'_{0,2} \left(\frac{y_1+y_2}{2}\right)\left(\frac{y'_1+y'_2}{2}\right)}, \
\eeqn
where
the normalization coefficients before and after this diagonalization are, of course, equal and satisfy
$\left(\alpha_1+C_{1,0}\right)\left(\alpha_1+C_{2,0}\right)=\frac{1}{2}\left(\alpha_1-\beta_1\right) \times
2\left[\alpha_1+\beta_1+2C_{2,0}\right]$
and
$\left(\alpha_2+C'_{0,1}\right)\left(\alpha_2+C'_{0,2}\right)=\frac{1}{2}\left(\alpha_2-\beta_2\right)
\times
2\left[\alpha_2+\beta_2+2C'_{0,2}\right]$.
Finally, one could equally work with the two-particle
Jacoby coordinates,
$\frac{x_1+x_2}{\sqrt{2}}$ and $\frac{x_1-x_2}{\sqrt{2}}$,
and the resulting physics would of course not change.

Just like for the reduced one-particle density matrices, see (\ref{SPECIES_1_and_2_1RDM_COEFFS}),
it turns out that the linear combinations of parameters
comprising the normalization constants are those which simplify the working equations.
Explicitly,
substituting (\ref{MIX_WF_DENS_PAR}) into (\ref{SPECIES1_12},\ref{SPECIES2_12})
one readily obtains
\beqn\label{SPECIES_1_and_2_2RDM_COEFFS}
& &
\alpha_1+\beta_1+2C_{2,0} =
m_1 \Omega_1 \frac{1}{1 + \frac{2}{N_1} \left( \frac{m_1N_1}{M} \frac{\Omega_1}{\omega} +
\frac{m_2N_2}{M} \frac{\Omega_1}{\Omega_{12}} - 1 \right)}, \nonumber \\
& &
\alpha_2+\beta_2+2C'_{0,2} =
m_2 \Omega_2 \frac{1}{1 + \frac{2}{N_2} \left( \frac{m_2N_2}{M} \frac{\Omega_2}{\omega} +
\frac{m_1N_1}{M} \frac{\Omega_2}{\Omega_{12}} - 1 \right)}. \
\eeqn
Together with the $\alpha_1+\beta_1$ and $\alpha_2+\beta_2$ combinations, see below,
these will govern the degree of pairs' fragmentation in the mixture.

The structure of (\ref{SPECIES1_SPECIES2_2_RDM_DIAG}) deserves discussion.
The reduced two-particle density matrices are diagonal in the relative coordinates of the pairs,
$x_1-x_2$ and $y_1-y_2$,
but are not yet diagonal in the center-of-mass
coordinates of the pairs, 
$\frac{x_1+x_2}{2}$ and $\frac{y_1+y_2}{2}$.
Fortunately, the latter can be diagonalized
with another usage of Mehler's formula.
This would end up with the natural pair functions of the mixture.
Thus, by comparing the respective structures of (\ref{SPECIES1_SPECIES2_2_RDM_DIAG}) and Mehler's formula,
one determines
the pairs' condensate fraction of species $1$ 
and that of species $2$,
\beqn\label{2RDMs_COND_FRAC_1_2}
& &
\!\!\!\!\!\!\!\!
1 - \rho^{(2)}_1 =
\frac{2\sqrt{\alpha_1+\beta_1+2C_{2,0}}}{\sqrt{\alpha_1+\beta_1} + \sqrt{\alpha_1+\beta_1+2C_{2,0}}} = \nonumber \\
& &
\!\!\!\!\!\!\!\!
= \frac{2}{\sqrt{\left[1 + \frac{2}{N_1} \left( \frac{m_1N_1}{M} \frac{\omega}{\Omega_1} + \frac{m_2N_2}{M} \frac{\Omega_{12}}{\Omega_1} - 1 \right)\right]\left[1 + \frac{2}{N_1} \left( \frac{m_1N_1}{M} \frac{\Omega_1}{\omega} +
\frac{m_2N_2}{M} \frac{\Omega_1}{\Omega_{12}} - 1 \right)\right]}+1}, \nonumber \\
& &
\!\!\!\!\!\!\!\!
1 - \rho^{(2)}_2 =
\frac{2\sqrt{\alpha_2+\beta_2+2C'_{0,2}}}{\sqrt{\alpha_2+\beta_2} + \sqrt{\alpha_2+\beta_2+2C'_{0,2}}} = \nonumber \\
& &
\!\!\!\!\!\!\!\!
= \frac{2}{\sqrt{\left[1 + \frac{2}{N_2} \left( \frac{m_2N_2}{M} \frac{\omega}{\Omega_2} + \frac{m_1N_1}{M} \frac{\Omega_{12}}{\Omega_2} - 1 \right)\right]\left[1 + \frac{2}{N_2} \left( \frac{m_2N_2}{M} \frac{\Omega_2}{\omega} +
\frac{m_1N_1}{M} \frac{\Omega_2}{\Omega_{12}} - 1 \right)\right]}+1}. \
\eeqn
Correspondingly,
the pairs' depleted fractions are
\beqn\label{2RDMs_DEPLT_FRAG_1_2}
& & \!\!\!\!\!\!\!\!
\rho^{(2)}_1 = \frac{\sqrt{\alpha_1+\beta_1} - \sqrt{\alpha_1+\beta_1+2C_{2,0}}}
{\sqrt{\alpha_1+\beta_1} + \sqrt{\alpha_1+\beta_1+2C_{2,0}}} = \nonumber \\
& & \!\!\!\!\!\!\!\!
= \frac{\sqrt{\left[1 + \frac{2}{N_1} \left( \frac{m_1N_1}{M} \frac{\omega}{\Omega_1} + \frac{m_2N_2}{M} \frac{\Omega_{12}}{\Omega_1} - 1 \right)\right]\left[1 + \frac{2}{N_1} \left( \frac{m_1N_1}{M} \frac{\Omega_1}{\omega} +
\frac{m_2N_2}{M} \frac{\Omega_1}{\Omega_{12}} - 1 \right)\right]}-1}{\sqrt{\left[1 + \frac{2}{N_1} \left( \frac{m_1N_1}{M} \frac{\omega}{\Omega_1} + \frac{m_2N_2}{M} \frac{\Omega_{12}}{\Omega_1} - 1 \right)\right]\left[1 + \frac{2}{N_1} \left( \frac{m_1N_1}{M} \frac{\Omega_1}{\omega} +
\frac{m_2N_2}{M} \frac{\Omega_1}{\Omega_{12}} - 1 \right)\right]}+1}, \nonumber \\
& & \!\!\!\!\!\!\!\!
\rho^{(2)}_2 = \frac{\sqrt{\alpha_2+\beta_2} - \sqrt{\alpha_2+\beta_2+2C'_{0,2}}}
{\sqrt{\alpha_2+\beta_2} + \sqrt{\alpha_2+\beta_2+2C'_{0,2}}} = \nonumber \\
& & \!\!\!\!\!\!\!\!
= \frac{\sqrt{\left[1 + \frac{2}{N_2} \left( \frac{m_2N_2}{M} \frac{\omega}{\Omega_2} + \frac{m_1N_1}{M} \frac{\Omega_{12}}{\Omega_2} - 1 \right)\right]\left[1 + \frac{2}{N_2} \left( \frac{m_2N_2}{M} \frac{\Omega_2}{\omega} +
\frac{m_1N_1}{M} \frac{\Omega_2}{\Omega_{12}} - 1 \right)\right]}-1}{\sqrt{\left[1 + \frac{2}{N_2} \left( \frac{m_2N_2}{M} \frac{\omega}{\Omega_2} + \frac{m_1N_1}{M} \frac{\Omega_{12}}{\Omega_2} - 1 \right)\right]\left[1 + \frac{2}{N_2} \left( \frac{m_2N_2}{M} \frac{\Omega_2}{\omega} +
\frac{m_1N_1}{M} \frac{\Omega_2}{\Omega_{12}} - 1 \right)\right]}+1}, \
\eeqn
and will be seen to govern the degree of fragmentation of pairs.
Combining the above,
the decomposition of the mixture's two-particle reduced
density matrices (\ref{SPECIES_1_and_2_2RDM}) 
is hence given by
\beqn\label{SPECIES1_SPECIES2_2_RDM_NAT_GEM}
& &
\rho_1^{(2)}(x_1,x_2,x'_1,x'_2) = 
N_1(N_1-1) \sum_{n_1=0}^\infty \left(1-\rho_1^{(2)}\right)\left(\rho_1^{(2)}\right)^{n_1}
\Phi^{(2)}_{1,n_1}(x_1,x_2) \Phi^{(2),\ast}_{1,n_1}(x'_1,x'_2), \nonumber \\
& & \Phi^{(2)}_{1,n_1}(x_1,x_2) =
\frac{1}{\sqrt{2^{n_1} {n_1}!}}
\left(\frac{s_1^{(2)}}{\pi}\right)^{\frac{1}{4}}
H_{n_1}\left[\sqrt{s_1^{(2)}}\left(\frac{x_1+x_2}{2}\right)\right]
e^{-\frac{1}{2}s_1^{(2)}\left(\frac{x_1+x_2}{2}\right)^2} \times \nonumber \\
& & \times
\left(\frac{m_1\Omega_1}{2\pi}\right)^{\frac{1}{4}}
e^{-\frac{1}{4}m_1\Omega_1\left(x_1-x_2\right)^2}, \nonumber \\
& &
\rho_2^{(2)}(y_1,y_2,y'_1,y'_2) = 
N_2(N_2-1) \sum_{n_2=0}^\infty \left(1-\rho_2^{(2)}\right)\left(\rho_2^{(2)}\right)^{n_2}
\Phi^{(2)}_{2,{n_2}}(y_1,y_2) \Phi^{(2),\ast}_{2,{n_2}}(y'_1,y'_2), \nonumber \\
& & \Phi^{(2)}_{2,{n_2}}(y_1,y_2) =
\frac{1}{\sqrt{2^{n_2} {n_2}!}}
\left(\frac{s_2^{(2)}}{\pi}\right)^{\frac{1}{4}}
H_{n_2}\left[\sqrt{s_2^{(2)}}\left(\frac{y_1+y_2}{2}\right)\right]
e^{-\frac{1}{2}s_2^{(2)}\left(\frac{y_1+y_2}{2}\right)^2} \times \nonumber \\
& & \times
\left(\frac{m_2\Omega_2}{2\pi}\right)^{\frac{1}{4}}
e^{-\frac{1}{4}m_2\Omega_2\left(y_1-y_2\right)^2}, \
\eeqn
where the scalings of the natural geminals are
\beqn\label{SPECIES1_SPECIES2_1_RDM_NAT_GEM_SCALINGS}
& &
\!\!\!\!\!\!\!\!\!\!\!\!\!\!\!\!
s^{(2)}_1 = 2\sqrt{\left(\alpha_1+\beta_1\right)\left(\alpha_1+\beta_1+2C_{2,0}\right)} =
2m_1\Omega_1
\sqrt{\frac{1 + \frac{2}{N_1} \left( \frac{m_1N_1}{M} \frac{\omega}{\Omega_1} + \frac{m_2N_2}{M} \frac{\Omega_{12}}{\Omega_1} - 1 \right)}{1 + \frac{2}{N_1} \left( \frac{m_1N_1}{M} \frac{\Omega_1}{\omega} +
\frac{m_2N_2}{M} \frac{\Omega_1}{\Omega_{12}} - 1 \right)}}, \nonumber \\
& &
\!\!\!\!\!\!\!\!\!\!\!\!\!\!\!\!
s^{(2)}_2 = 2\sqrt{\left(\alpha_2+\beta_2\right)\left(\alpha_2+\beta_2+2C'_{0,2}\right)} =
2m_2\Omega_2
\sqrt{\frac{1 + \frac{2}{N_2} \left( \frac{m_2N_2}{M} \frac{\omega}{\Omega_2} + \frac{m_1N_1}{M} \frac{\Omega_{12}}{\Omega_2} - 1 \right)}{1 + \frac{2}{N_2} \left( \frac{m_2N_2}{M} \frac{\Omega_2}{\omega} +
\frac{m_1N_1}{M} \frac{\Omega_2}{\Omega_{12}} - 1 \right)}}. \
\eeqn
Hence, we obtain in closed form,
as a function of all parameters of the imbalanced mixture,
the pairs' condensate fractions, the pairs' depleted fractions and fragmentation (see below),
as well as the natural geminals of
species $1$ and $2$ bosons.

The pairs' depletions $\rho_1^{(2)}$ and $\rho_2^{(2)}$ determine the degree of pair fragmentation of each species,
where the respective occupations 
$\left(1-\rho_1^{(2)}\right)\left(\rho_1^{(2)}\right)^{n_1}, n_1=0,1,2,\ldots$
and
$\left(1-\rho_2^{(2)}\right)\left(\rho_2^{(2)}\right)^{n_2}, n_2=0,1,2,\ldots$
have the same algebraic form just like the natural occupation numbers.
A close inspection and comparison of the natural orbitals' and geminals' occupations
tells one that, in a given mixture,
the degree of pairs' fragmentation is
always slightly larger than
the degree of bosons' fragmentation,
see more in the illustrative example below.
 
\section{Application}\label{APPLICATION}

\begin{figure}[!]
\begin{center}
\hglue -1.2 truecm
\includegraphics[width=0.36\columnwidth,angle=-90]{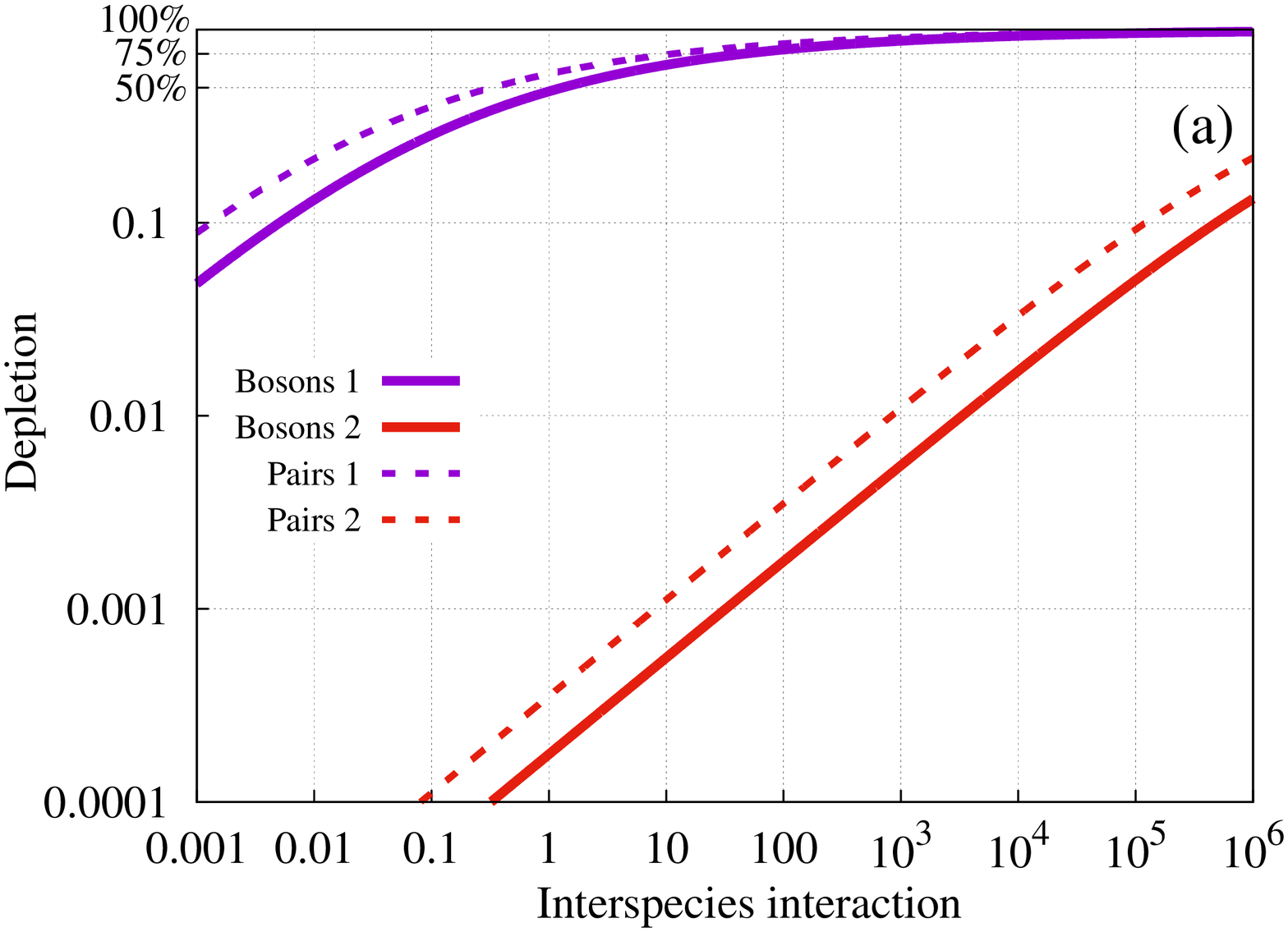}
\hglue -0.25 truecm
\includegraphics[width=0.36\columnwidth,angle=-90]{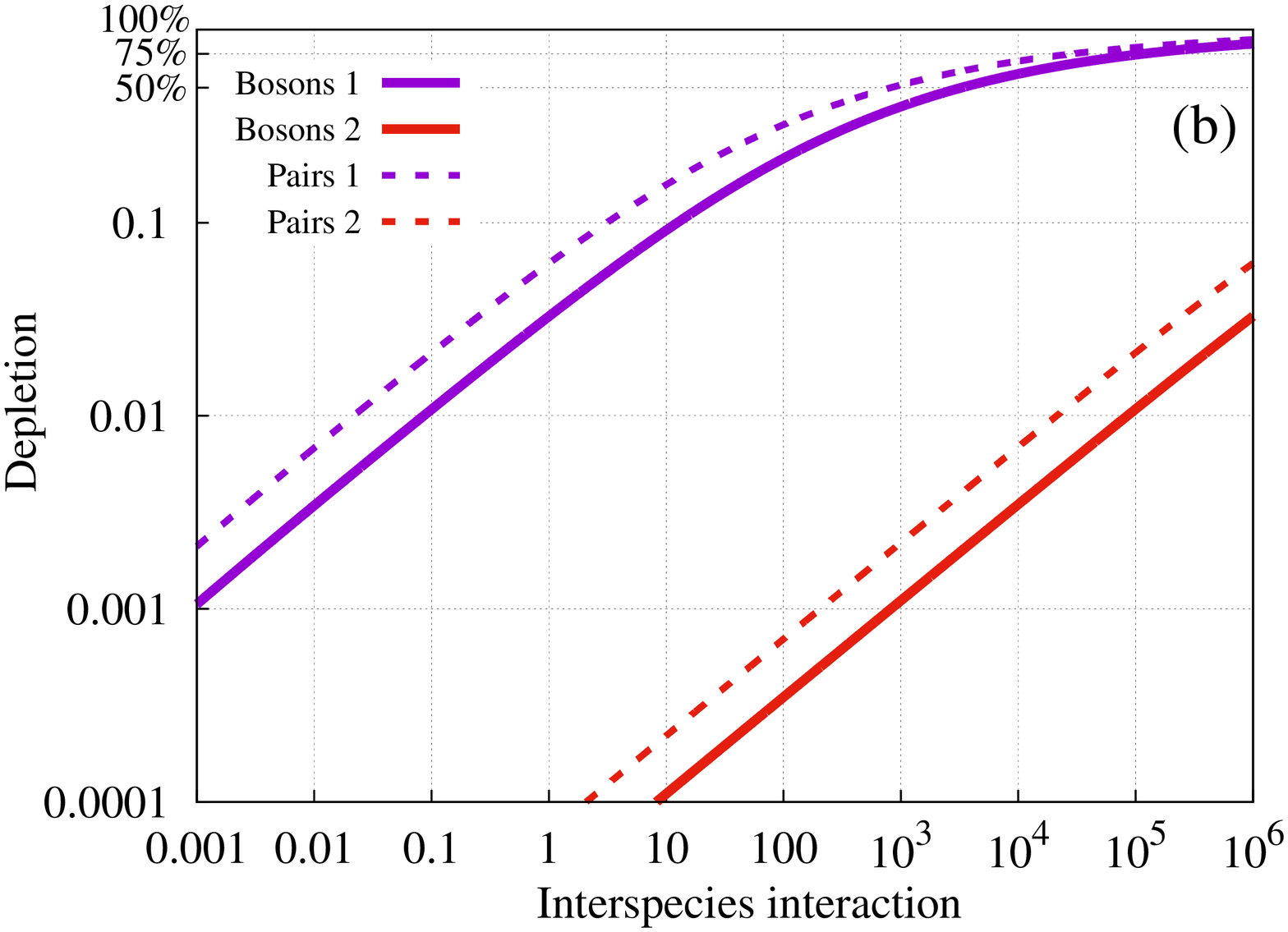}
\hglue -1.2 truecm
\includegraphics[width=0.36\columnwidth,angle=-90]{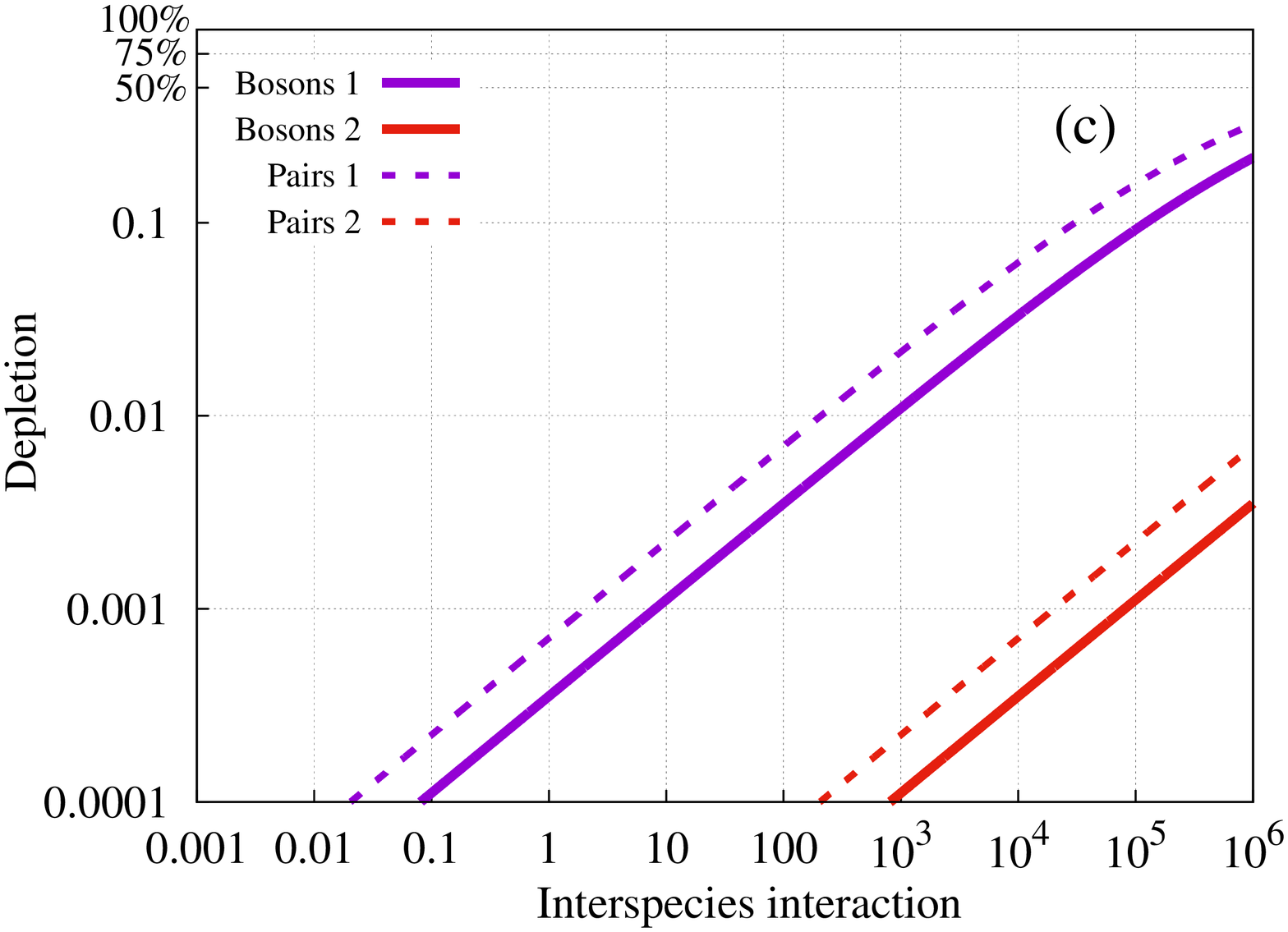}
\hglue -0.25 truecm
\includegraphics[width=0.36\columnwidth,angle=-90]{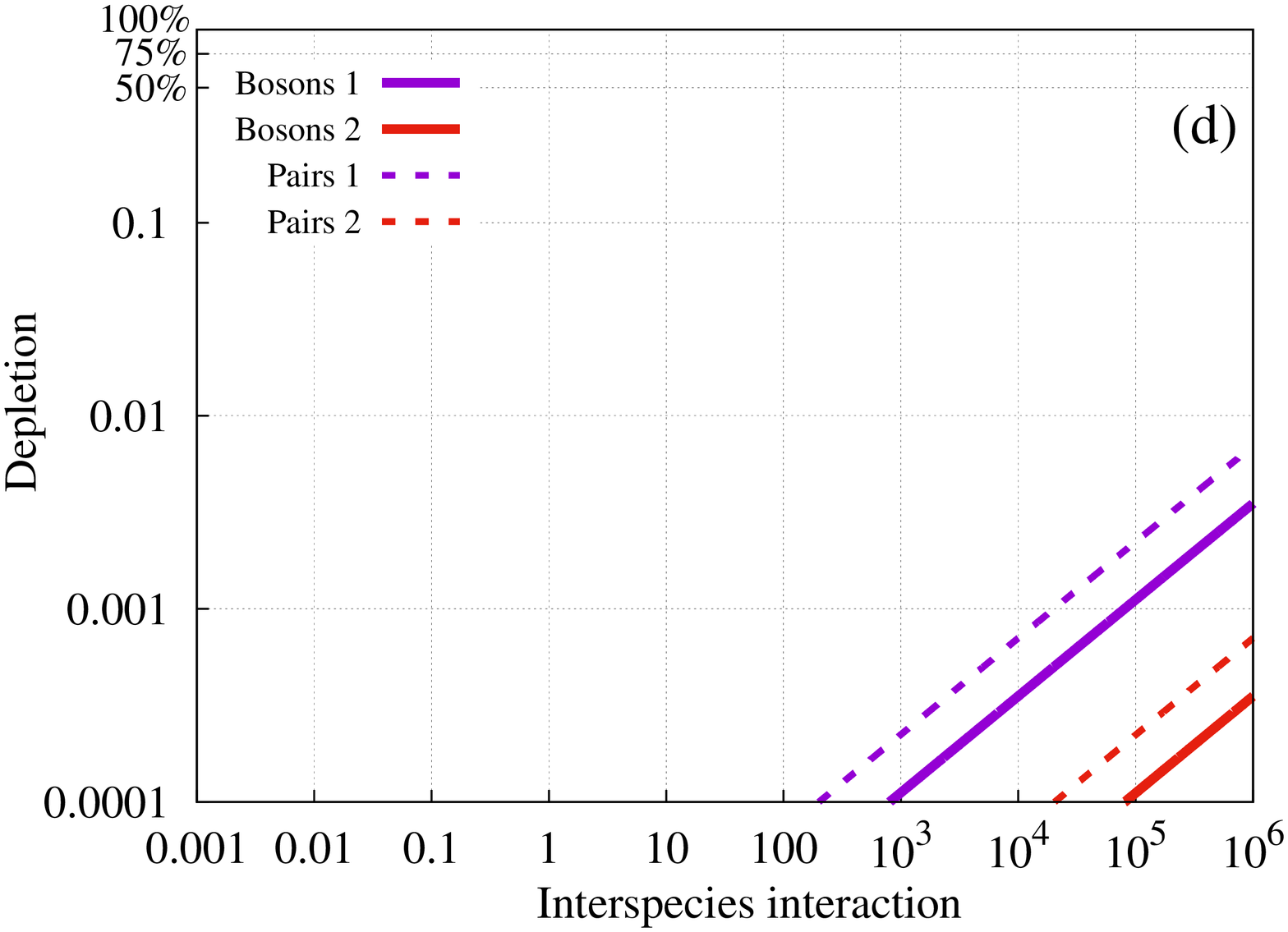}
\end{center}
\vglue 0.5 truecm
\caption{Depletion and fragmentation of bosons and pairs in a mixture made of
$N_1=100$ species $1$ bosons (the system) and $N_2=10^6$ species $2$ bosons (the bath)
as a function of the interspecies interaction.
The intraspecies interactions of species $1$ and $2$ bosons vanish.
Effects of the mass ratio: $m_1=1$ and (a) $m_2=\frac{1}{10,000}$, (b) $m_2=\frac{1}{100}$, (c) $m_2=1$, and (d) $m_2=100$. 
The lighter the bosons comprising the bath are the stronger
is the fragmentation of the system's bosons and pairs.
The curves correspond to the analytically-computed depletions
(from which fragmentation is given) 
of the reduced one-particle and two-particle density matrices
given in Eqs.~(\ref{1RDMs_DEPLT_FRAG_1_2}) and (\ref{2RDMs_DEPLT_FRAG_1_2}), respectively.  
See the text for more details.
The quantities shown are dimensionless.}
\label{F1}
\end{figure}

\begin{figure}[!]
\begin{center}
\hglue -1.2 truecm
\includegraphics[width=0.36\columnwidth,angle=-90]{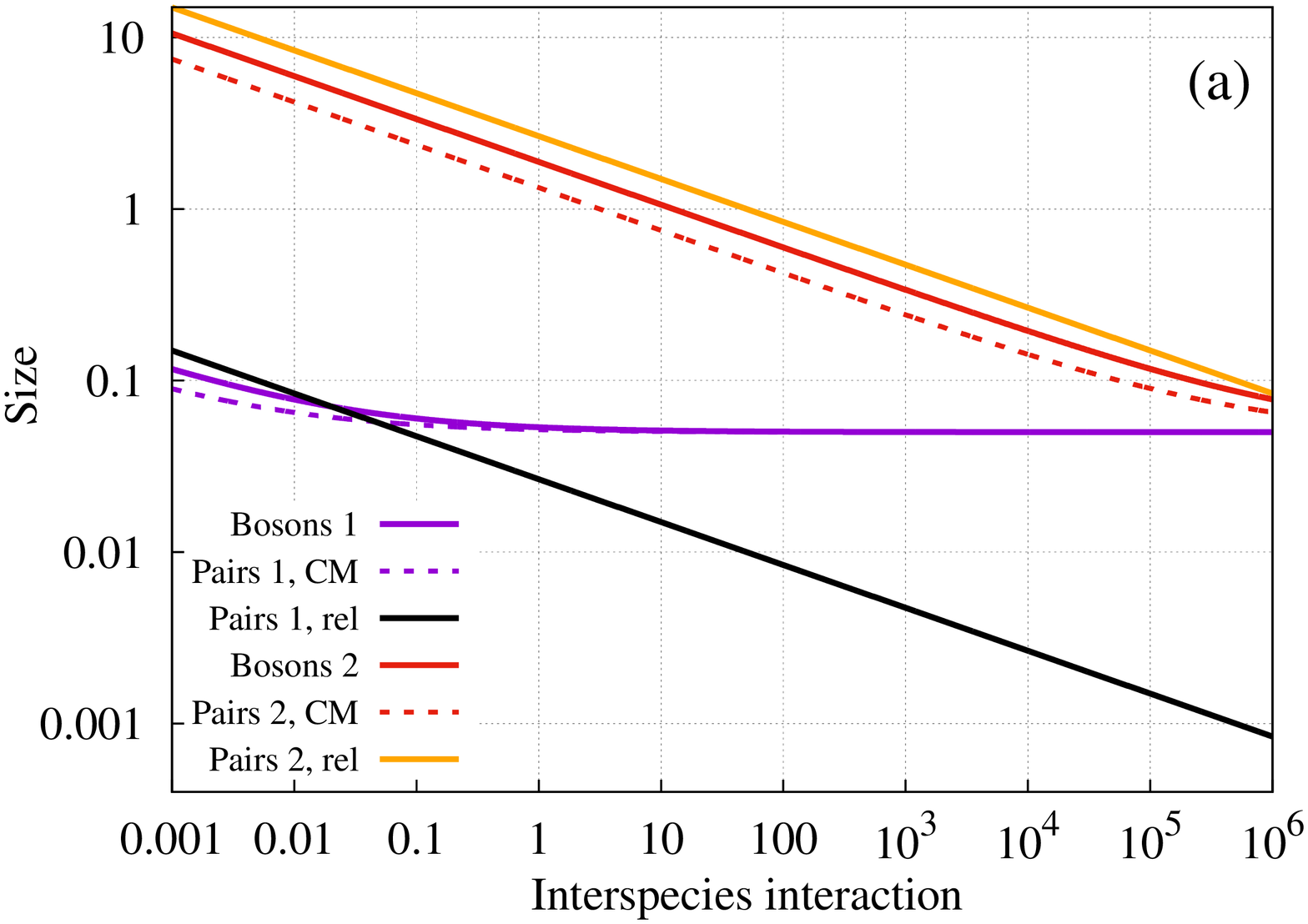}
\hglue -0.25 truecm
\includegraphics[width=0.36\columnwidth,angle=-90]{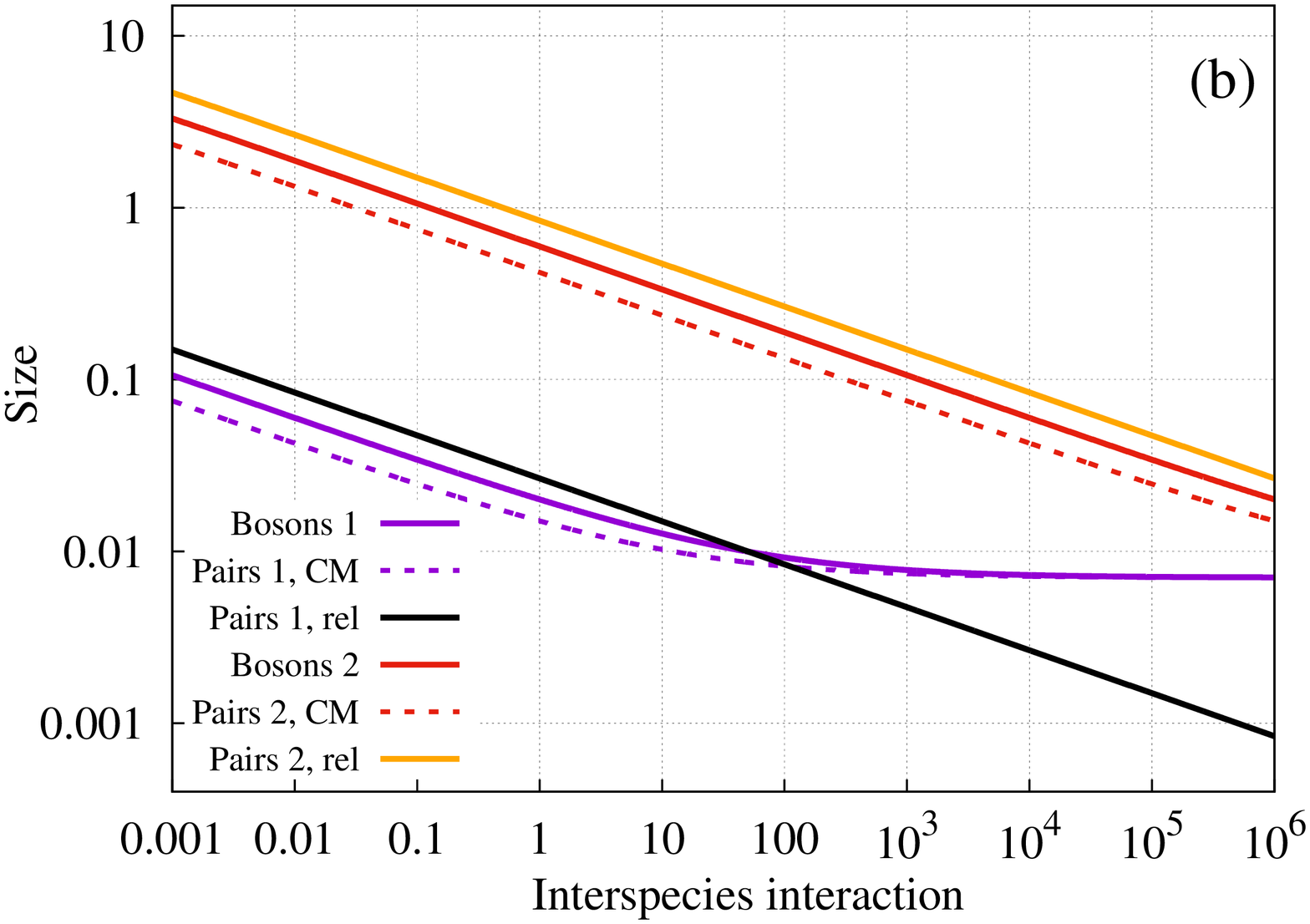}
\hglue -1.2 truecm
\includegraphics[width=0.36\columnwidth,angle=-90]{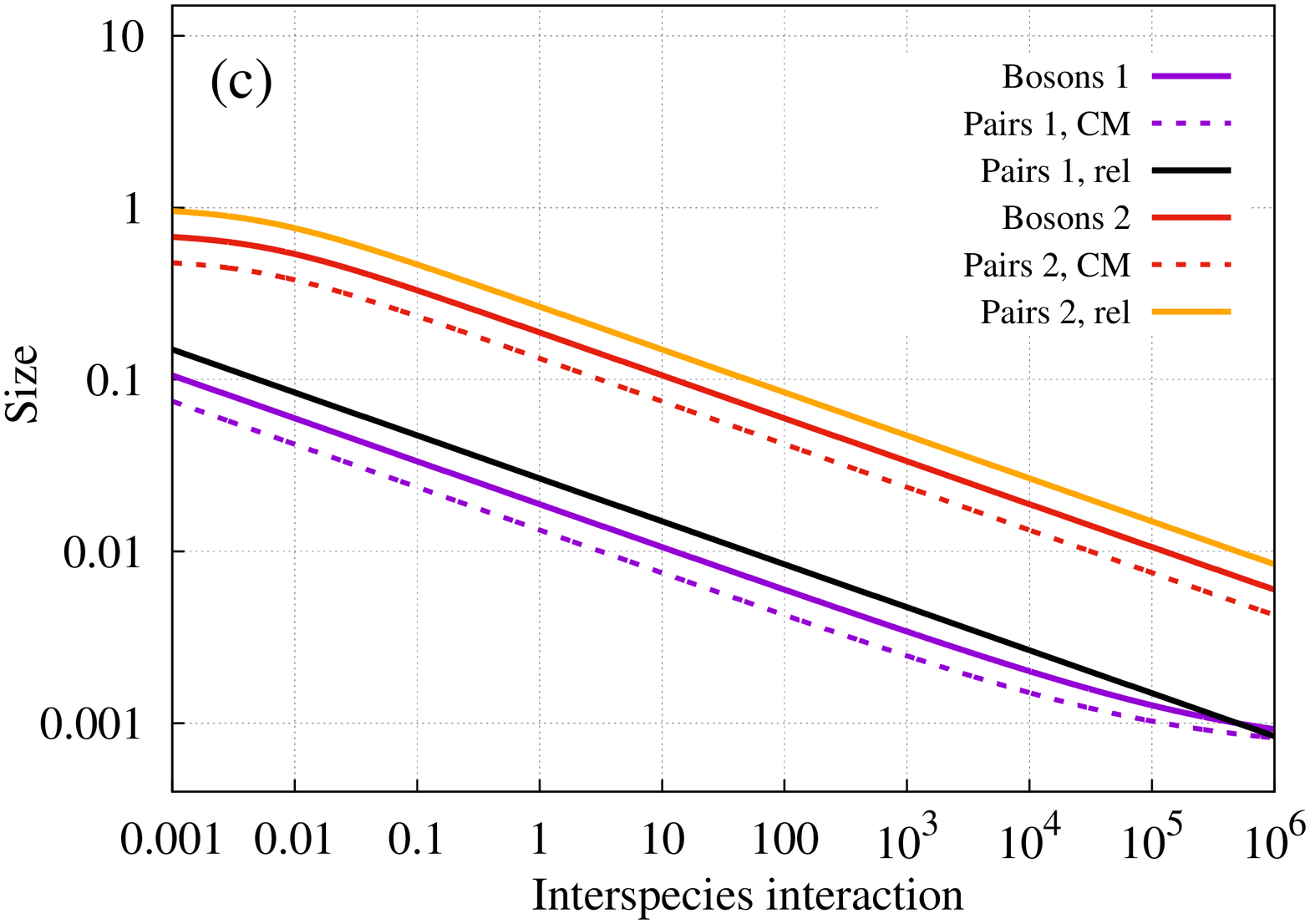}
\hglue -0.25 truecm
\includegraphics[width=0.36\columnwidth,angle=-90]{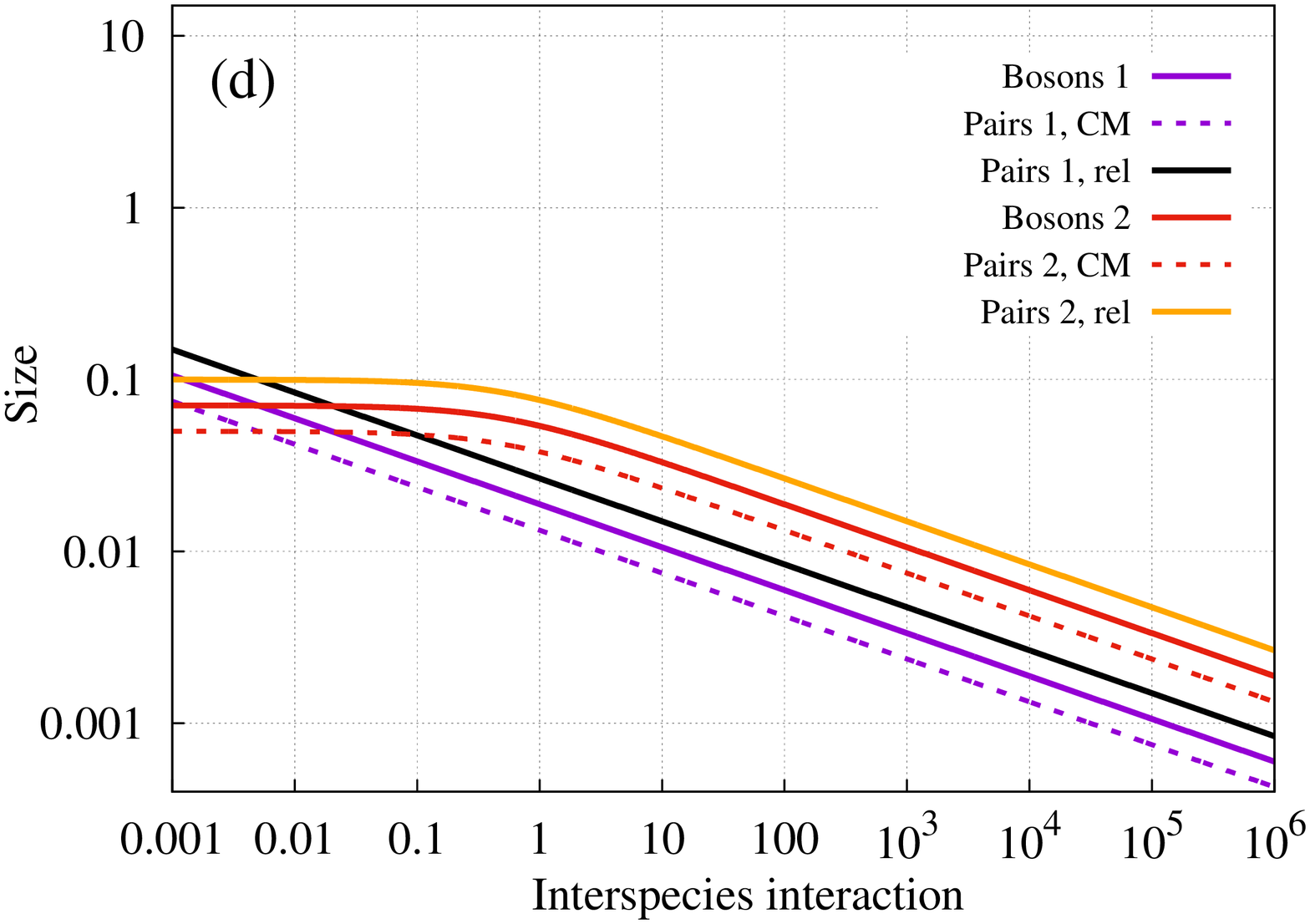}
\end{center}
\vglue 0.5 truecm
\caption{Size of bosons and pairs in a mixture made of
$N_1=100$ species $1$ bosons (the system) and $N_2=10^6$ species $2$ bosons (the bath)
as a function of the interspecies interaction.
The intraspecies interactions of species $1$ and species $2$ bosons vanish.
Effects of the mass ratio: $m_1=1$ and (a) $m_2=\frac{1}{10,000}$, (b) $m_2=\frac{1}{100}$, (c) $m_2=1$, and (d) $m_2=100$. 
The lighter the bosons comprising the bath are the larger is the size of the system's bosons and pairs.
The curves correspond to the analytically-computed
widths of the respective one-particle and two-particle densities.
See Fig.~\ref{F1} for the respective fragmentations
and the text for more details.
The quantities shown are dimensionless.}
\label{F2}
\end{figure}

Consider a mixture made of $N_1=100$ species $1$ bosons (the system)
embedded in $N=10^6$ species $2$ bosons (the bath).
Let there be no intraspecies interactions, i.e., $\lambda_1=0$ and $\lambda_2=0$.
The only interaction is interspecies which we consider below to be attractive,
namely, $\lambda_{12} > 0$.
The mixture's frequencies
$\Omega_{1} = \sqrt{\omega^2 + \frac{2N_2}{m_1}\lambda_{12}}$ and
$\Omega_{2} = \sqrt{\omega^2 + \frac{2N_1}{m_2}\lambda_{12}}$
simplify and
$\Omega_{12} = \sqrt{\omega^2 + 2\left(\frac{N_1}{m_2}+\frac{N_2}{m_1}\right)\lambda_{12}}$
remains of course unchanged.
The questions we are interested in are whether the interaction between
the species can lead to fragmentation of the otherwise non-interacting system and bath,
and how to manipulate the fragmentation.
We fix the trap's frequency as $\omega=1$.
The only degree of freedom left is the mass ratio of species $1$ and $2$ bosons.
Accordingly, we fix the mass of the system's boson to $m_1=1$
and vary the mass of the bath's boson such that
$m_2 = \frac{1}{10,000}$, $m_2 = \frac{1}{100}$,
$m_2 = 1$, or $m_2 = 100$.
Figures \ref{F1} and \ref{F2} collect the results.

In Fig.~\ref{F1}, the depletion and hence the fragmentation of each of the species is depicted.
Several trends are observed:
(i) Increasing the interspecies attraction leads to increase in the depletion and fragmentation
of the system and the bath;
(ii) The larger bath is always less depleted than the smaller system;
(iii) Pairs, of either species, are always slightly more fragmented than the bosons of the same species;
(iv) The lighter the mass of the bath's bosons is the stronger are the depletion and fragmentation of the system.
Since the trapping potential is fixed, 
only the interspecies interaction and masses determine the size of
system and bath.
Hence, we proceed along this direction 
in search for further properties of 
the fragmented mixture.

The one-particle density of species $1$ is
$\rho_1^{(1)}(x) =
N_1 \left(\frac{\alpha_1+C_{1,0}}{\pi}\right)^{\frac{1}{2}} e^{-\left(\alpha_1+C_{1,0}\right)x^2}$
and that of species $2$ is
$\rho_2^{(1)}(y) =
N_2 \left(\frac{\alpha_2+C'_{0,1}}{\pi}\right)^{\frac{1}{2}} e^{-\left(\alpha_2+C'_{0,1}\right)y^2}$.
The two-particle densities of species $1$ and $2$ are
$\rho_1^{(2)}(x_1,x_2) = 
N_1(N_1-1)
\left(\frac{\alpha_1-\beta_1}{2\pi}\right)^{\frac{1}{2}}
e^{-\frac{1}{2}\left(\alpha_1-\beta_1\right)\left(x_1-x_2\right)^2}
\left(\frac{2\left[\alpha_1+\beta_1+2C_{2,0}\right]}{\pi}\right)^{\frac{1}{2}} \times$
$\times e^{-2\left[\alpha_1+\beta_1+2C_{2,0}\right]\left(\frac{x_1+x_2}{2}\right)^2}$
and
$\rho_2^{(2)}(y_1,y_2) = 
N_2(N_2-1)
\left(\frac{\alpha_2-\beta_2}{2\pi}\right)^{\frac{1}{2}}
e^{-\frac{1}{2}\left(\alpha_2-\beta_2\right)\left(y_1-y_2\right)^2} \times$
$\times \left(\frac{2\left[\alpha_2+\beta_2+2C'_{0,2}\right]}{\pi}\right)^{\frac{1}{2}}
e^{-2\left[\alpha_2+\beta_2+2C'_{0,2}\right]\left(\frac{y_1+y_2}{2}\right)^2}$,
where all coefficients have been prescribed and simplified above.
Then, we can readily compute the sizes of the bosons' clouds,
$\sigma_{1}^{(1)} = \sqrt{\frac{1 + \frac{1}{N_1} \left( \frac{m_1N_1}{M} \frac{\Omega_1}{\omega} +
\frac{m_2N_2}{M} \frac{\Omega_1}{\Omega_{12}} - 1 \right)}{2m_1 \Omega_1}}$
and
$\sigma_{2}^{(1)} = \sqrt{\frac{1 + \frac{1}{N_2} \left( \frac{m_2N_2}{M} \frac{\Omega_2}{\omega} +
\frac{m_1N_1}{M} \frac{\Omega_2}{\Omega_{12}} - 1 \right)}{2m_2 \Omega_2}}$,
and investigate how they depend explicitly on all parameters and especially
the interaction between the system and the bath.
Figure \ref{F2} collects the results.

There are two degrees-of-freedom for a pair of indistinguishable bosons (say of species $1$),
the center-of-mass $\frac{x_1+x_2}{2}$ and relative $x_1-x_2$ coordinates
which are symboled {\it CM} and {\it rel} hereafter.
Thus, we have for the sizes of the pairs' clouds
$\sigma_{1,CM}^{(2)} = \sqrt{\frac{1 + \frac{2}{N_1} \left( \frac{m_1N_1}{M} \frac{\Omega_1}{\omega} +
\frac{m_2N_2}{M} \frac{\Omega_1}{\Omega_{12}} - 1 \right)}{4m_1 \Omega_1}}$
and
$\sigma_{1,rel}^{(2)} = \sqrt{\frac{1}{m_1 \Omega_1}}$ for species $1$
and
$\sigma_{2,CM}^{(2)} = \sqrt{\frac{1 + \frac{2}{N_2} \left( \frac{m_2N_2}{M} \frac{\Omega_2}{\omega} +
\frac{m_1N_1}{M} \frac{\Omega_2}{\Omega_{12}} - 1 \right)}{4m_2 \Omega_2}}$
and
$\sigma_{2,rel}^{(2)} = \sqrt{\frac{1}{m_2 \Omega_2}}$ for species $2$.
The sizes of bosons and pairs are monotonously decreasing functions,
in conjunction with the attractive interspecies interaction, see Fig.~\ref{F2}.
Side by side,
for strong interspecies attraction we find saturation of the sizes of the bosons' clouds for both species,
$\lim_{\lambda_{12} \to \infty} \sigma_{1}^{(1)} = \lim_{\lambda_{12} \to \infty} \sigma_{2}^{(1)}
= \sqrt{\frac{1}{2M}}$
and saturation of the sizes of the center-of-mass coordinates of the pairs' clouds for both species,
$\lim_{\lambda_{12} \to \infty} \sigma_{1,CM}^{(2)} = \lim_{\lambda_{12} \to \infty} \sigma_{2,CM}^{(2)}
= \sqrt{\frac{1}{2M}}$.
This limiting size of bosons and pairs is the same,
and is dictated solely by the inverse mass of the whole mixture.
For the parameters of the illustrated example,
these saturations are already seen for the system's bosons and pairs in Fig.~\ref{F2}a,b.
Hence, size saturation
goes hand in hand with the strong depletion and fragmentation, see Fig.~\ref{F1}a,b.
Finally,
the sizes of the relative coordinates of the pairs' clouds of both species vanish,
$\lim_{\lambda_{12} \to \infty} \sigma_{1,rel}^{(2)} = \lim_{\lambda_{12} \to \infty} \sigma_{2,rel}^{(2)} = 0$,
indicating the effective strong attraction between each pair of indistinguishable bosons
resulting from the strong interspecies interaction.
This is a good location to conclude the present investigation.

\section{Summary}\label{SUMMARY}

We have presented an analytical and concise many-body investigation,
utilizing a solvable many-particle model,
on fragmentation in a trapped imbalanced bosonic mixture.
First, we derived closed-form expressions for the eigenvalues and eigenfunctions
of the reduced one-particle and two-particle density matrices of the mixture.
These expressions depict in a transparent manner the
dependence of the fragmentation of each species
on the masses, numbers of bosons,
and the intraspecies and interspecies interactions.

We then picked up one application,
in which a bosonic system made of $N_1=100$ particles
and a bosonic bath made of $N_2=10^6$ particles are
both made of non-interacting bosons,
and investigated the emergence of fragmentation
in the system due to its interaction with the bath.
We find fragmentation of
bosons and pairs in the bath,
which becomes stronger when the mass of the bath's bosons is decreased.
The size of bosons and pairs is computed and analyzed.

There are many other applications that come to mind,
of which we mention just three.
We have investigated a scenario with non-interacting system and bath bosons.
It would be instructive to investigate how the properties of
two fragmented species could be altered by the interspecies interaction between them.
Here, either of the three interactions may be attractive or repulsive,
adding another facet of interest and possibilities.
We have explored fragmentation of pairs of identical
bosons in the mixture.
Recently, the fragmentation of indistinguishable
pairs in the specific and simpler case of a balanced mixture was studied \cite{Atoms_2021}.
It would be rewarding to develop the appropriate tools 
and explore the fragmentation of
indistinguishable pairs in the imbalanced mixture.
Last but not least,
we foresee that extending the present analytical many-body investigation
of statics properties to time-dependent scenarios is a path worth following.

\ack

This research was supported by the Israel Science Foundation (Grant No. 1516/19). 

\appendix

\section{Working expressions for the reduced one-particle and two-particle density matrices}\label{APPENDIX}

The coefficients in the reduced one-particle and two-particle density matrices,
Eqs.~(\ref{SPECIES_1_and_2_1RDM}) and (\ref{SPECIES_1_and_2_2RDM}),
are obtained 
from recursive relations following the reduction of 
the all-particle density matrix \cite{JPA_2017,Floquet_2020} and read
\beqn\label{SPECIES1_12}
& & \!\!\!\!\!\!\!\!\!\!
\alpha_1 + C_{1,0} = (\alpha_1-\beta_1)
\frac{[(\alpha_1-\beta_1)+N_1\beta_1][(\alpha_2-\beta_2)+N_2\beta_2]-\gamma^2N_1N_2}
{[(\alpha_1-\beta_1)+(N_1-1)\beta_1][(\alpha_2-\beta_2)+N_2\beta_2]-\gamma^2(N_1-1)N_2}, \\
& & \!\!\!\!\!\!\!\!\!\!
\alpha_1 + \beta_1 + 2C_{2,0} = (\alpha_1-\beta_1)
\frac{[(\alpha_1-\beta_1)+N_1\beta_1][(\alpha_2-\beta_2)+N_2\beta_2]-\gamma^2N_1N_2}
{[(\alpha_1-\beta_1)+(N_1-2)\beta_1][(\alpha_2-\beta_2)+N_2\beta_2]-\gamma^2(N_1-2)N_2} \nonumber \
\eeqn
for species $1$ bosons and
\beqn\label{SPECIES2_12}
& & \!\!\!\!\!\!\!\!\!\!
\alpha_2 + C'_{0,1} = (\alpha_2-\beta_2)
\frac{[(\alpha_1-\beta_1)+N_1\beta_1][(\alpha_2-\beta_2)+N_2\beta_2]-\gamma^2N_1N_2}
{[(\alpha_2-\beta_2)+(N_2-1)\beta_2][(\alpha_1-\beta_1)+N_1\beta_1]-\gamma^2(N_2-1)N_1}, \\
& & \!\!\!\!\!\!\!\!\!\!
\alpha_2 + \beta_2 + 2C'_{0,2} = (\alpha_2-\beta_2)
\frac{[(\alpha_1-\beta_1)+N_1\beta_1][(\alpha_2-\beta_2)+N_2\beta_2]-\gamma^2N_1N_2}
{[(\alpha_2-\beta_2)+(N_2-2)\beta_2][(\alpha_1-\beta_1)+N_1\beta_1]-\gamma^2(N_2-2)N_1} \nonumber \
\eeqn
for species $2$ bosons.
Therein,
the parameters of the all-particle density matrix (\ref{N_DENS}) are 
\beqn\label{MIX_WF_DENS_PAR}
& & \alpha_1 = m_1\Omega_1 + \beta_1 =
m_1 \Omega_1 \left[1 + \frac{1}{N_1} \left( \frac{m_1N_1}{M} \frac{\omega}{\Omega_1} + \frac{m_2N_2}{M} \frac{\Omega_{12}}{\Omega_1} - 1 \right)\right], \\
& & \beta_1 = m_1 \Omega_1 \frac{1}{N_1} \left( \frac{m_1N_1}{M} \frac{\omega}{\Omega_1} + \frac{m_2N_2}{M} \frac{\Omega_{12}}{\Omega_1} - 1 \right), \nonumber \\
& & \alpha_2 = m_2\Omega_2 + \beta_2 =
m_2 \Omega_2 \left[1 + \frac{1}{N_2} \left( \frac{m_2N_2}{M} \frac{\omega}{\Omega_2} + \frac{m_1N_1}{M} \frac{\Omega_{12}}{\Omega_2} - 1 \right)\right], \nonumber \\
& & \beta_2 = m_2 \Omega_2 \frac{1}{N_2} \left( \frac{m_2N_2}{M} \frac{\omega}{\Omega_2} + \frac{m_1N_1}{M} \frac{\Omega_{12}}{\Omega_2} - 1 \right), \nonumber \\
& & \gamma = \frac{m_1m_2}{M}(\Omega_{12}-\omega) =
m_1 \Omega_1 \frac{1}{N_2} \frac{m_2N_2}{M}\left( \frac{\Omega_{12}}{\Omega_1} - \frac{\omega}{\Omega_1} \right) =
m_2 \Omega_2 \frac{1}{N_1} \frac{m_1N_1}{M}\left( \frac{\Omega_{12}}{\Omega_2} - \frac{\omega}{\Omega_2} \right). \nonumber \ 
\eeqn
Equations (\ref{SPECIES_1_and_2_1RDM_COEFFS}) and (\ref{SPECIES_1_and_2_2RDM_COEFFS}) of the main text
are obtained by substituting the expressions for $\alpha_1$, $\alpha_2$, $\beta_1$, $\beta_2$, and $\gamma$
in (\ref{MIX_WF_DENS_PAR})
into Eqs.~(\ref{SPECIES1_12}) and (\ref{SPECIES2_12}).

\end{document}